\documentclass[amsmath,twocolumn,prb,aps]{revtex4-1}
\usepackage{amsthm,amsfonts,graphicx,verbatim, color}
\usepackage{bm}
\usepackage[utf8]{inputenc}

\newcommand{\psq}{\ensuremath{|\vec p|^2}}
\newcommand{\pisq}{\ensuremath{|\vec \Pi|^2}}
\newcommand{\EE}{\ensuremath{\beta}}
\newcommand{\alphacrit}{\alpha_{\text{crit.}}}

\begin{document}

\title{Connection between Fermi contours of zero-field electrons and $\nu=\frac12$ composite fermions in two-dimensional systems}

\author{Matteo Ippoliti, Scott D. Geraedts, and R. N. Bhatt}

\affiliation{Departments of Electrical Engineering and Physics, Princeton University, Princeton NJ 08544, USA}

\begin{abstract}
We investigate the relation between the Fermi sea (FS) of zero-field carriers in two-dimensional systems and the FS of the corresponding composite fermions which emerge in a high magnetic field at filling $\nu = \frac{1}{2}$, as the kinetic energy dispersion is varied.
We study cases both with and without rotational symmetry, and find that there is generally no straightforward relation between the geometric shapes and topologies of the two FSs.
In particular, we show analytically that the composite Fermi liquid (CFL) is completely insensitive to a wide range of changes to the zero-field dispersion which preserve rotational symmetry, including ones that break the zero-field FS into multiple disconnected pieces.
In the absence of rotational symmetry, we show that the notion of `valley pseudospin' in many-valley systems is generically not transferred to the CFL, in agreement with experimental observations.
We also discuss how a rotationally symmetric band structure can induce a reordering of the Landau levels,  
opening interesting possibilities of observing higher-Landau-level physics in the high-field regime.
\end{abstract}

\maketitle

\section{Introduction}

In recent years, the issue of geometry of quantum Hall states has attracted
significant attention. Starting with the formulation by Haldane \cite{Haldane2011}, several
investigations, first numerical \cite{BoYang2012, Wang2012, Papic2013} and then experimental \cite{Kamburov2013, Kamburov2014} as well as theoretical \cite{Yang2013, Murthy2013, Balram2016}, have been exploring the role
of anisotropy in fractional quantum Hall (FQH) states \cite{BoYang2012, Papic2013, Wang2012,Kamburov2013, Kamburov2014} as well the effect of metric deformations \cite{Johri2016}.
Mass anisotropy in the zero field electron dispersion is naturally present in electron-doped many-valley semiconductors such as AlAs \cite{Shkolnikov2002}, which makes the Fermi contours at $B = 0$ elliptical rather than circular. 
Another way to generate anisotropy in the FQH regime is application of an in-plane magnetic field\cite{Lilly1999}.

For non-interacting electrons, this problem can be made isotropic by a rescaling of coordinates ({\it i.e.}, by a shear deformation). 
However, as emphasized by Haldane \cite{Haldane2011},
anisotropy of the mass and that of the dielectric tensor (which determines electron-electron interactions) in electronic materials are, in general, quite different. 
Thus, for fractional quantum Hall states where electron-electron interactions are essential, the anisotropy cannot be eliminated by simple rescaling of coordinates as in the non-interacting case.
The distinct anisotropies of the electron mass and dielectric tensor lead to an extra
degree of freedom in Laughlin's variational scheme for fractional Hall state: 
the optimal coordinate rescaling parameter for the fractional state. 
This was demonstrated in variational as well as numerical studies of the anisotropic problem \cite{BoYang2012}, and later generalized to the case of tilted magnetic fields, where rotational symmetry is also broken \cite{Papic2013}.

While the early numerical studies \cite{BoYang2012, Wang2012} concentrated on the gapped states, in particular
the primary Laughlin $\nu = 1/3$ state, a comparison with experimental transport data
was complicated by two facts:
firstly, the gaps are somewhat dependent on anisotropy, 
and secondly, and more importantly, conductivity involves the anisotropy of the transport relaxation time in addition to that of the mass. 
However, exquisite experiments \cite{Kamburov2013,Kamburov2014}
performed on the gapless $\nu = \frac12$ state directly measured the composite fermion (CF)
Fermi contour anisotropy
\footnote{The Fermi contour is the boundary of the Fermi sea}, 
and showed that distortions of the zero-field band Fermi contour from a circular shape resulted in a corresponding distortion of the CF Fermi contour at magnetic fields corresponding to
half-filling of the lowest Landau level. 
The latest measurements on samples subject to uniaxial strain\cite{Jo2017}
show clearly that an elliptical distortion of the Fermi contour at $B = 0$ leads to an elliptical distortion of the Fermi contour for the CF Fermi liquid state at half filling, 
albeit with a with reduced anisotropy, which is in excellent agreement with numerical studies of
electron systems with a parabolic dispersion but anisotropic mass\cite{Ippoliti1}. 
Moreover, experimental studies of hole systems with warping of the zero field Fermi contour from circular symmetry to square ({\it i.e.}, discrete four-fold rotational) symmetry show transference of this warping to the CF Fermi contour in a tilted magnetic field\cite{Mueed2015}.
Numerical investigations of this problem find a weak, but nonzero response of the CFL Fermi contour to band deformations with four-fold rotational symmetry\cite{Ippoliti2}.

These observations might lead one to conclude that there is an intimate connection between the Fermi contour of composite fermions at $\nu = \frac12$ and the original zero-field Fermi contour.
A fundamental question of interest is thus the following: 
can the $B=0$ electronic band structure be used to engineer different phases in the high field, Landau level regime? 
If this approach proved successful in significantly altering not only single-particle eigenvalues, but also eigenfunctions (which often determine the phase for interacting systems), it would in effect lead to Landau-level engineering, parallel to the enormously successful zero-field band-structure engineering in traditional semiconductor physics.

In this paper we show that this expectation is too optimistic, and provide examples that demonstrate how the connection between the zero-field and high-field regimes is more subtle, indeed tenuous in some instances.
In Section \ref{sec:exact}, we consider dispersions with circular symmetry, a case which can be solved exactly. 
We discuss a number of cases that display a wide variety of zero-field Fermi seas, but nevertheless share the {\it same} Fermi sea for CFs at high magnetic fields.
In Section \ref{sec:anisotropic} we generalize our study to a case with anisotropy (two-fold discrete rotational symmetry) where the zero-field Fermi sea has two {\it disconnected} Fermi pockets, and demonstrate that the Fermi sea of composite Fermions, while retaining the two-fold rotational symmetry, remains {\it fully connected}. 
This suggests that the only memory of the zero-field Fermi contour that is retained in the high field limit is the nature of its rotational symmetry, while much of the other details are wiped out.

\section{Exact results for circularly symmetric dispersions}
\label{sec:exact}

In this section, we consider the Landau level problem for a single, non-degenerate
electron band with a circularly symmetric, but otherwise arbitrary, electron dispersion at $B =0$.
We show that the problem in the high field limit reduces to the canonical Landau
problem with extensively degenerate Landau bands with the {\it same} eigenfunctions as
the standard parabolic one, but with eigenvalues that generally differ from the canonical
harmonic oscillator values. As a result, in the high magnetic field limit, the problem
reduces to that of a single Landau level; the only issue that remains is – which
Landau level has the lowest eigenvalue? If it is the $N=0$ Landau level, we find that the
interacting problem for fractional filling is {\it exactly the same} as that for the parabolic
case, and consequently the CF Fermi sea is completely
unchanged for a whole range of zero field electron dispersions, including cases
where the $B = 0$ Fermi sea changes from being a circle, to being an annulus, 
and also cases where the Fermi sea has multiple disconnected pieces. In short, while the zero
field Fermi sea undergoes Lifshitz transitions, the high field problem remains
unaffected. 
This demonstrates that the state at high field is completely unresponsive to rotationally symmetric distortions of the zero-field problem, showing that the relation between the two problems can be subtle. 

Specifically, we consider a zero-field Hamiltonian with an arbitrary, but isotropic kinetic energy term in two dimensions of the form:
\begin{equation}
H_0=E_0 f(\psq /2mE_0)
\label{H0}
\end{equation}
where $m$ is the free electron mass and $E_0$ is an energy scale, which makes the argument of $f$ dimensionless. 
For non-relativistic free fermions, one has $f(x)=x$. 
Note that the Hamiltonian depends on $\vec p$ only through $\psq = p_x^2 + p_y^2$. 
In order for the problem to be well defined, we require that $f(x)$ be analytic and bounded from below, but set no other constraints on it.
In the presence of a magnetic field $B$ in the $z$-direction, we have $ \vec p \to \vec{\Pi} \equiv \vec p - e\vec A$, where $\vec \nabla \times \vec A = B$.

We note in passing that non-parabolic dispersions could occur in any strongly interacting fermionic system with continuous translational and full rotational symmetry ({\it i.e.} Galilean invariance). In fact, significant deviations have been predicted and seen experimentally in Galilean invariant, rotationally symmetric three dimensional dilute mixtures of $^3$He in $^4$He\cite{Varma, Bhatt} which depend on the effective interaction psuedopotentials\cite{Hsu}. 

In the `usual' Landau problem with a quadratically dispersing Hamiltonian, the eigenstates $\{ |\phi_N \rangle \}$ are clearly eigenstates of the operator $|\vec \Pi|^2$, with eigenvalues 
 \begin{equation}
\epsilon_N^{\rm quadratic} = (N + 1/2) E_c = (N+ 1/2)\EE E_0.
\label{EVLL}
\end{equation}
Here N is the Landau level index, $E_c = heB/m$ is the cyclotron energy, and $\EE$ is the ratio of $E_c$ to $E_0$.
Since the Hamiltonian in Eq.~(\ref{H0}) is a function of $|\vec \Pi|^2$ only, 
the $\{ |\phi_N \rangle \}$ are eigenstates of Eq.~(\ref{H0}) for any $f(x)$, and therefore even $f(x)$ which dramatically change the zero-field Fermi surface don't affect Fermi surface in the high field problem (as long as 
$|\phi_0\rangle$, the lowest Landau level in the case of quadratic dispersion, still has the lowest energy). 
These energies are given by:
\begin{equation}
H_0 | \phi_N \rangle = \epsilon_N |\phi_N\rangle,
\quad
\epsilon_N=E_0 f( [N+1/2]\EE)
\label{energy}
\end{equation}

Before proceeding further, we establish conventions that we follow in the rest of the paper.
We use as independent variables the Landau level filling fraction $\nu$, and either the zero-field Fermi energy $\epsilon_F$ or the electron density $n_e$, depending on convenience. For a given dispersion, fixing $\nu$ and $\epsilon_F$ (or $\nu$ and $n_e$) fixes the magnetic field $B$, and hence the cyclotron energy.
If we define $\mathcal A(\epsilon_F)$ as the area in $k$-space occupied by the Fermi sea at given chemical potential $\epsilon_F$, we find that the ratio of the cyclotron energy to $E_0$ is
\begin{equation}
\EE \equiv \frac{E_c}{E_0} = \frac{\mathcal A(\epsilon_F)k_0^{-2} }{2\pi \nu}.
\label{eq:cyclotron}
\end{equation} 
where 
\begin{equation}
k_0 \equiv \frac{\sqrt{E_0m}}{\hbar}\;.
\end{equation}

is a characteristic wavevector.

We now discuss the simplest non-trivial example for $f(x)$:
\begin{equation}
\begin{aligned}
f(x) & =   -C x +4x^2  , \\
\mathcal H_0(C) & = -C\frac{\pisq}{2m}+\frac{1}{E_0}\left(\frac{\pisq}{m} \right)^2 \, .
\label{f1eqn}
\end{aligned}
\end{equation}
Here $C$ is a dimensionless parameter.
This kinetic energy leads to a Fermi sea which is an annulus if $C>0$ and $\epsilon_F<0$, and a circle otherwise. 
The annular case can be seen as a simplified model for a realistic energy dispersion that has recently been observed \cite{Jo2017B} for holes in GaAs quantum wells.

In order to demonstrate the lack of correspondence between the zero-field and composite fermion Fermi contours at magnetic field corresponding to a half-filled lowest Landau level ($\nu = \frac{1}{2}$), it suffices to show a pair of values $(C,\epsilon_F)$ where the zero-field Fermi sea is annular and the $N=0$ Landau level has the lowest energy.
In that case, the single-particle orbitals in the high field (lowest eigenvalue) limit are the same as in the usual parabolic case ({\it i.e.}, $N=0$ Landau level wavefunctions). Thus the CF Fermi sea is a circle, whereas the electron Fermi sea is an annulus.  

\begin{figure}
\includegraphics[width=0.97\linewidth]{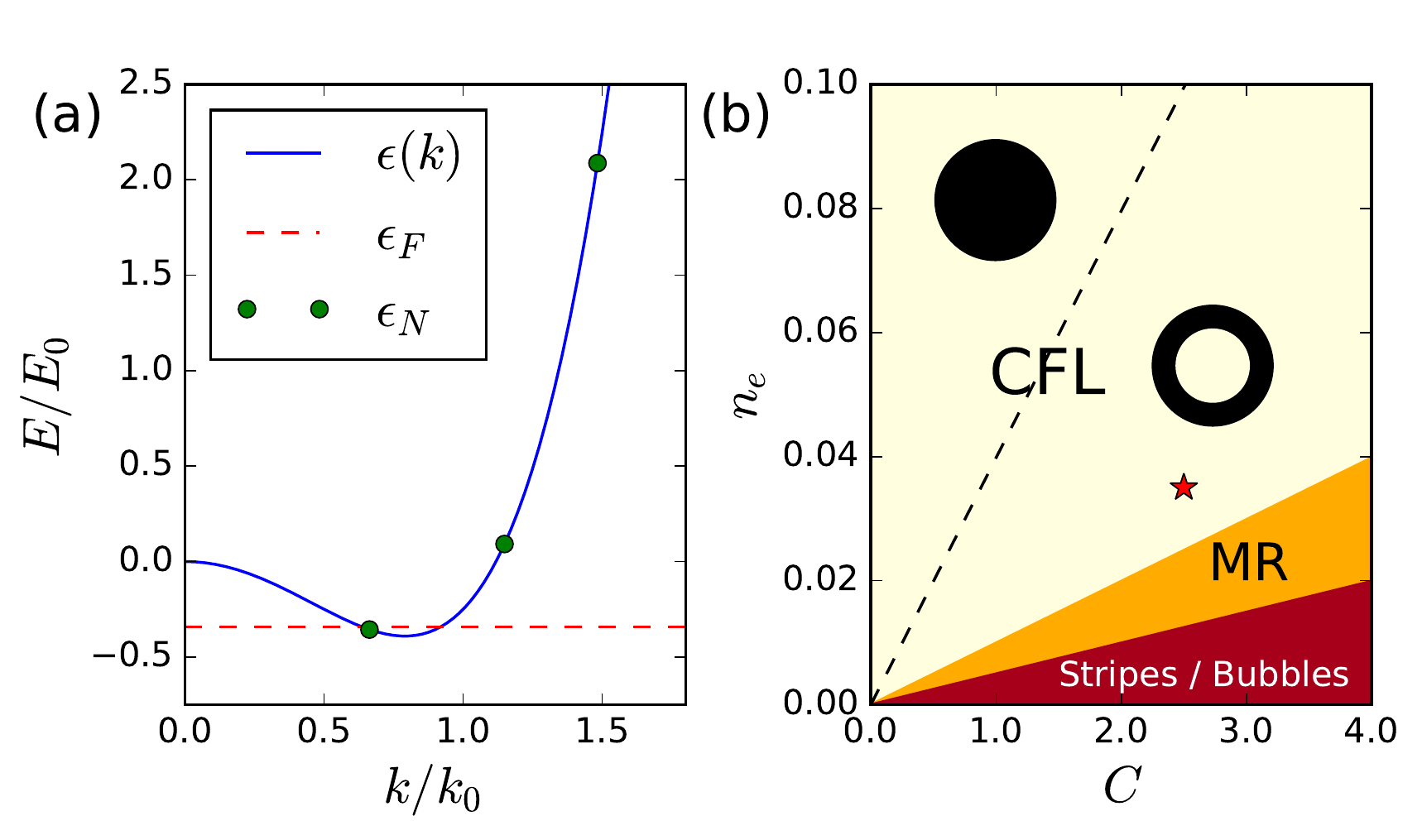}
\caption{
(a) Dispersion of Hamiltonian \eqref{f1eqn} with $C=2.5$ (solid line), Fermi energy $\epsilon_F=-0.34 E_0$ (dashed line) and energies of the lowest three Landau levels (dots). 
The $B=0$ Fermi sea (consisting of $k$-states below the dashed line) is an annulus. 
At these values of $C$ and $\epsilon_F$, the $N=0$ Landau level has the lowest energy, so the ground state is a CFL with a circular Fermi sea.
(b) A plot of the phase diagram of the system described by Eq.~(\ref{f1eqn}), as a function of $C$ and electron density $n_e$ (expressed in units of $k_0^2$), in a magnetic field tuned to half-filling. 
Depending on which $N$ minimizes $\epsilon_N$, the system can be in a CFL phase ($N=0$), a Moore-Read phase ($N=1$), or a stripe or bubble phase ($N\geq 2$).
The dashed line in the CFL phase corresponds to $\epsilon_F=0$, i.e. to the transition from circular to annular electron Fermi seas at $B=0$.
The star denotes the parameters used in (a).
}
\label{f1}
\end{figure}

We found such a combination of parameters to be quite generic.
Fig.~\ref{f1}(a) shows an example of the band structure of a system with $C=2.5$ and $\epsilon_F=-0.34 E_0$.
The system clearly has an annular Fermi sea and the $N=0$ Landau level is the one with lowest energy. 
Fig.~\ref{f1}(b) shows the phase diagram as a function of $C$ and the electron density 
\begin{equation}
n_e = \frac{k_0^2}{2\pi} \sqrt{(C/4)^2+\epsilon_F/E_0}\;,
\end{equation}
displaying a large region of parameter space where these conditions are met.
At smaller $C$ or larger density, the zero-field Fermi sea becomes a circle (though the CFL is completely insensitive to this transition).

Figure \ref{f1}(b) also contains other phases. For a general function $f(x)$, there are a number of different possibilities depending on the values of $N \equiv \text{argmin}_{N'}(\epsilon_{N'})$:
\paragraph*{i.}
If $N=0$, the ground state of the interacting system will be a composite Fermi liquid with a {\it circular} Fermi contour, despite the fact that the zero-field non-interacting Fermi contour may have a more complicated (albeit circularly symmetric) shape.
\paragraph*{ii.}
If $N = 1$, the interacting system will be in the Moore-Read phase, with ground state described either by the Pfaffian or anti-Pfaffian model wavefunctions (which one is chosen depends generically on the effects of Landau level mixing, which is affected by the particular choice of $f(x)$).
\paragraph*{iii.}
If $N \geq 2$, we generically expect the ground state at every filling $0<\nu<1$ (not just the value $\nu = \frac{1}{2}$ used in the figure) to be in a stripe or bubble phase.
This opens exciting possibilities for studies of such phases in the high-field limit, a regime opposite to the one in which they are normally observed.

We can see that the latter case is realized even in the simple example of a Eq.~(\ref{f1eqn}) at large enough $C$ and small enough density.
For example, we find that these phases are predicted to occur at a realistic carrier density $\sim 10^{11} {\rm cm}^{-2}$ for band parameters $C \sim 3$ and $E_0 \sim 3 {\rm meV}$.
The recently observed annular Fermi sea for holes in GaAs quantum wells\cite{Jo2017B} would most likely not give rise to these phases, due to the fact that the annular Fermi pocket is included in a larger Fermi sea ({\it i.e.} there is no gap between the ``interesting'' band and other bands).
On the other hand, band structures that are reasonably well approximated by Eq.~(\ref{f1eqn}) are expected to occur quite generically in band-inverted semiconductors, such as HgTe, when a gap is opened {\it e.g.} by application of strain\cite{Moon2006}.

The above analysis is sufficient to demonstrate that there is not necessarily a relationship between zero-field and $\nu=\frac{1}{2}$ composite fermion Fermi contours. 
This analysis could be easily repeated for the case where a cubic term is added to Eq.~(\ref{f1eqn}).
The resulting dispersion, for appropriate values of the coefficients, describes isotropic dispersions with a roton-like minimum that were conjectured by Pitaevskii for dilute mixtures of $^3$He in $^4$He (see {\it e.g.} Ref. \onlinecite{Varma, Bhatt, Hsu}).
For appropriate values of the parameters, such dispersion could even lead to a Fermi sea with two disconnected pieces (a circle around $k = 0$ and an annulus between two larger values of $k$). 
For electron bands in solids, such non-monotonic dispersions with multiple Fermi pockets in zero magnetic field are quite common; 
however, because of the crystal structure, they possess discrete rotational symmetry, unlike the continuous symmetry considered here.
Numerical investigations, however, show that this type of anisotropy has a small effect on the resulting quantum Hall physics\cite{Ippoliti2}, so that the main effect is still expected to be the reordering of Landau levels induced by non-monotonicity of the dispersion.

Finally, in order to show that our conclusion is completely generic, in Appendix~\ref{app:rings} we consider an example of $f(x)$ which allows us to generate a zero-field Fermi sea consisting of an arbitrary number of disconnected components, while at the same time having $\text{argmin}_N(\epsilon_N) =  0$ when the magnetic field is such that $\nu = \frac{1}{2}$ (see Figure~\ref{fig:target}).

\section{Breaking rotational symmetry: systems with multiple Fermi pockets}
\label{sec:anisotropic}
\begin{figure}
\centering
\includegraphics[width = \columnwidth]{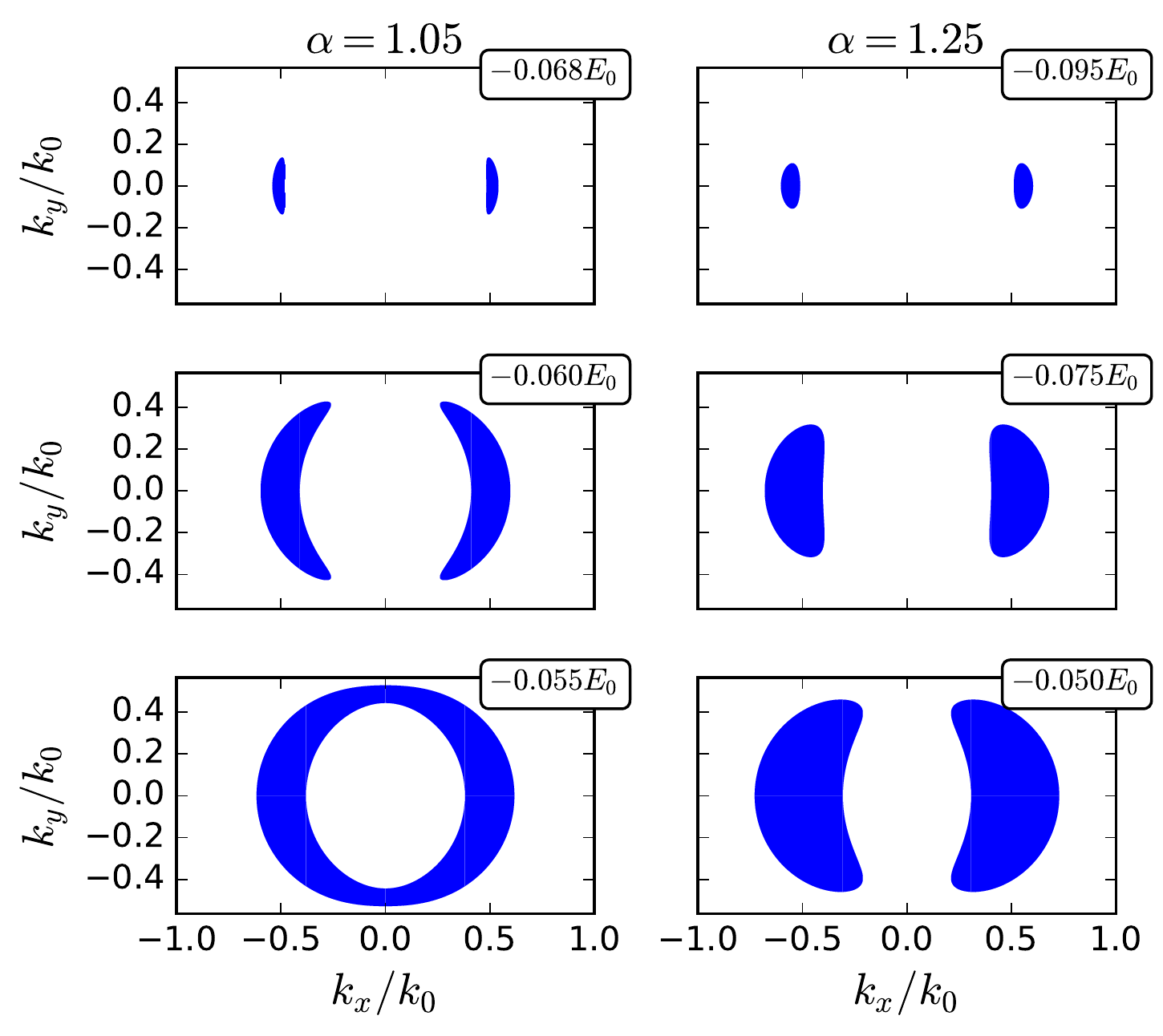}
\caption{Examples of zero-field Fermi seas obtained from the kinetic energy Eq.~\eqref{H_anis} at small values of the anisotropy parameter, $\alpha = 1.05$ (left) and $\alpha = 1.25$ (right), at various Fermi energies shown in the upper right corner of each panel.
Tuning these parameters allows continuous interpolation between the annular shape discussed in Section~\ref{sec:exact} and the Fermi pockets discussed in Section~\ref{sec:anisotropic}.}
\label{fig:bananas}
\end{figure}

The previous section shows that one can have a Fermi sea made of multiple disconnected pieces at zero field, and yet only a single composite Fermi sea at high magnetic field. 
We can observe a similar phenomenon for Fermi seas consisting of disconnected ``pockets'' without rotational symmetry, such as those shown in Fig.~\ref{fig:bananas}. 
These Fermi contours were generated from the following zero-field Hamiltonian:
\begin{equation}
H_0(\alpha) = -\left( \frac{\alpha p_x^2}{2m} +\frac{p_y^2}{2m\alpha}\right) 
+\frac{1}{E_0}\left(\frac{p_x^2+p_y^2}{m} \right)^2\;.
\label{H_anis}
\end{equation}
This is the same as Eq.~(\ref{f1eqn}) with $C=1$, except that we have included an anisotropy parameter $\alpha$ in the quadratic part which explicitly breaks rotational symmetry\cite{Ippoliti1}. 

\begin{figure}
\includegraphics[width=0.97\linewidth]{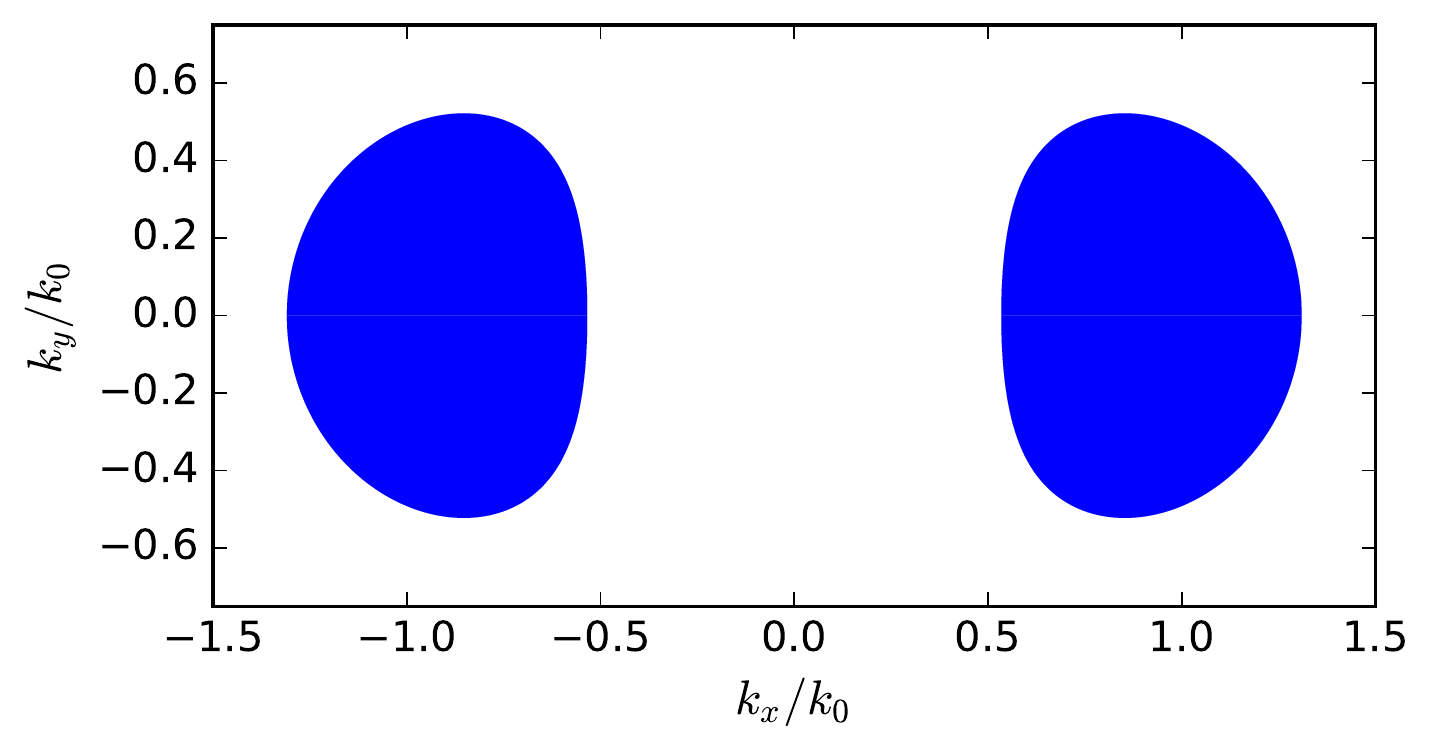}
\caption{Zero-field Fermi sea for the Hamitonian in Eq.~(\ref{H_anis}), using $\alpha=4$ and the Fermi energy $\epsilon_F=-0.49E_0$.
For these parameters, the inverse magnetic length at filling $\nu = \frac{1}{2}$ is $\ell_B^{-1} = 0.64 k_0$,
so the two Fermi pockets are separated by $\approx 2\ell_B^{-1}$.}
\label{fig:beans}
\end{figure}

Fermi contours with shapes similar to those in Fig.~\ref{fig:bananas} have been observed in GaAs systems in a parallel field\cite{Kamburov2014}, as well as in the surface states of Sn$_{1-x}$Pb${_x}$Se\cite{SnSpectra} and bismuth\cite{BismuthSpectra, Yazdani2016} (though in the latter case the valleys are elongated radially, rather than tangentially).
In such systems it is tempting to assume that multiple zero-field Fermi pockets imply that the system can be treated as having multiple valleys.
Here we show that this assumption is not always correct; 
even a Fermi sea like that of Fig.~\ref{fig:beans}, made of clearly separated pockets, does not necessarily imply two nearly degenerate low-lying Landau levels.

\begin{figure}
\includegraphics[width=0.97\linewidth]{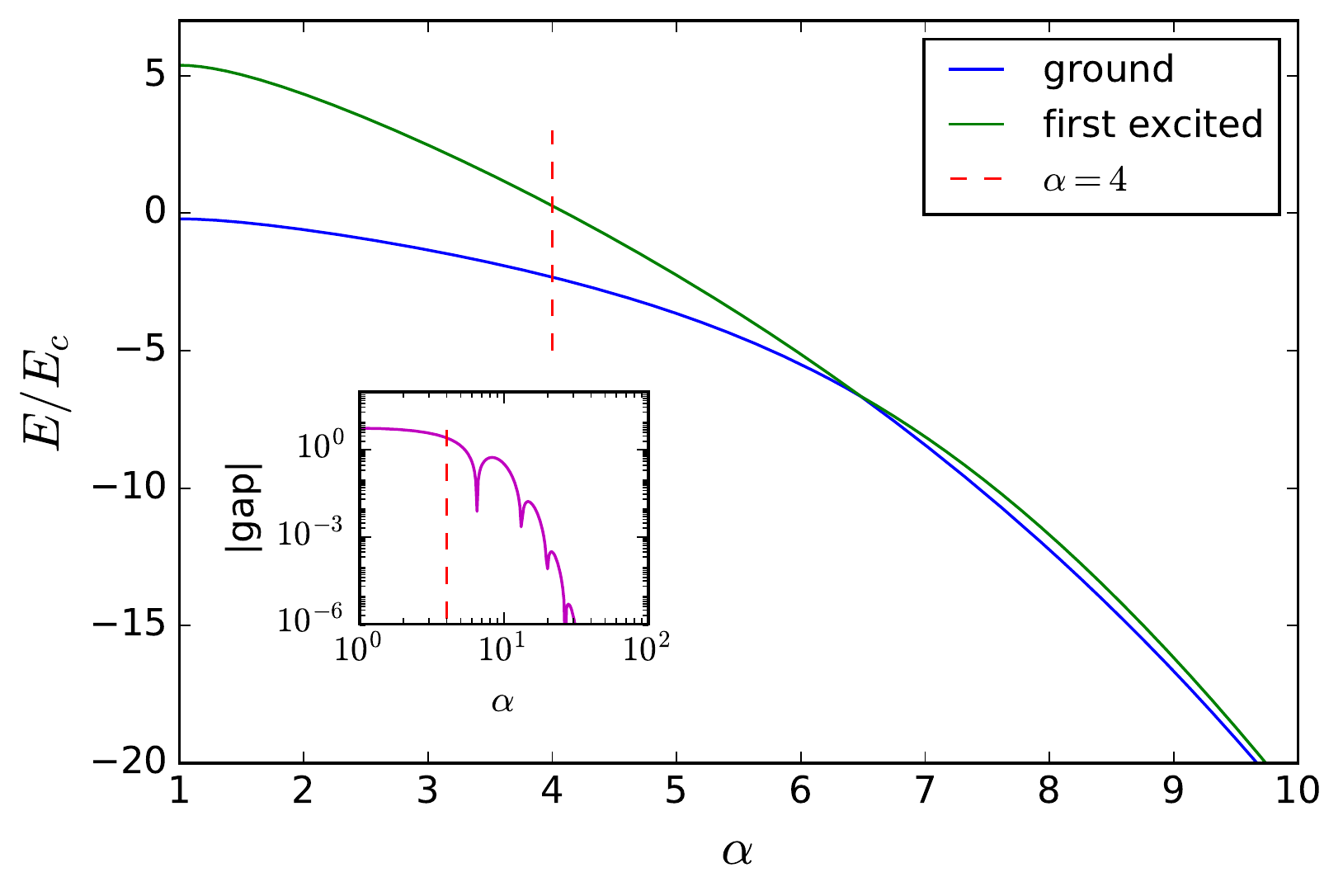}
\caption{The two lowest energies (in units if the cyclotron energy $E_c = \beta E_0$) of the Hamiltonian in Eq.~(\ref{H_anis}) placed in a magnetic field such that the electron density of Fig.~\ref{fig:beans} corresponds to filling $\nu = \frac{1}{2}$.
As the $\alpha$ parameter changes, the chemical potential is changed so that electron density and magnetic field stay constant.
Inset: difference between the two energies.
The dashed lines indicate $\alpha=4$, the value used in Fig.~\ref{fig:beans}.
Even with a Fermi sea made of two well-separated pockets, there is still a substantial energy gap between the lowest-lying Landau levels (roughly half of that at the isotropic point).
}
\label{fig:energies}
\end{figure}

To prove this point, for the dispersion Eq.~(\ref{H_anis}), we choose the Fermi energy $\epsilon_F$ such that there are two disconnected zero-field pockets (as in Fig.~\ref{fig:beans}). Since we work at $\nu = \frac{1}{2} $, the Fermi energy also sets the cyclotron energy.
We solve the single-particle Schrodinger equation using Eq.~(\ref{H_anis}) in the strong magnetic field determined by the Fermi energy and the filling fraction. 
Since the single-particle problem is symmetric upon interchanging the valleys, the orbitals will be equal superpositions of the two valleys. 
Naively this seems to conflict with the notion of having states in well-separated valleys, as all the electrons are in both valleys at once. 
The resolution to this is that if the two lowest-energy solutions are very close in energy (compared to the interaction strength), interactions can hybridize the states and lead to electron orbits which are purely in one valley or the other\cite{Sodemann2017}. 
It would be a mistake, however, to assume that this hybridization always happens when the zero-field Fermi surface has multiple pockets. If the splitting between the two lowest-lying single-particle solutions is much larger than the strength of interactions, $\sim e^2/\ell_B \varepsilon$, then this hybridization will not occur and there is no meaningful notion of `valley pseudospin' in the problem; 
rather, only one generalized Landau level matters. 
The `valley pseudospin' degree of freedom emerges only when the valleys are very well separated in units of $\ell_B^{-1}$ (the typical spread in momentum space of a Landau orbital).

Fig.~\ref{fig:energies} shows the energies of the two lowest solutions to the single-particle problem as a function of $\alpha$. The value $\alpha=4$, which was used to generate the Fermi contours in Fig.~\ref{fig:beans}, is indicated by the dashed line. 
We see that, as one turns $\alpha$ up from the isotropic $\alpha=1$ point, a large splitting between the two energies persists long after the problem has developed two disconnected Fermi contours. 
This energy splitting should be compared to the interaction energy to determine whether the system develops valleys. In quantum Hall problems we often assume that the interaction energy is much smaller than the cyclotron energy, i.e. the Landau level mixing parameter $\kappa$ (the ratio of interaction energy to cyclotron energy) is $\ll 1$. 
We can see that at large $\alpha$, even a small $\kappa$ will be enough to hybridize the levels and give well-defined valleys (e.g. from the plot we see that at $\alpha=10$, $\kappa\approx 10^{-3}$ would be sufficient). 
But up to $\alpha \sim 4$, the energy gap is still substantial, and a value of $\kappa \approx 1$ would be required to effectively hybridize the Landau orbitals.

To make this analysis concrete, we performed infinite density-matrix renormalization group\cite{Zaletel2013} calculations on the system described by Eq.~(\ref{H_anis}) placed on an infinite cylinder of circumference $L = 13 \ell_B$. 
Our simulation contains the lowest two Landau levels, whose single-particle energies are shown in Fig.~\ref{fig:energies}.
One of the Landau levels is symmetric under $k \mapsto -k$, the other is anti-symmetric; 
we label them by 0 and 1 respectively (even though their single-particle energies are not necessarily in that order).
The two Landau levels have different form factors which are determined by the eigenvectors corresponding to the energies shown in the figure. 
With this data, we can obtain the many-body ground state at total filling $\nu = \frac{1}{2}$ and evaluate the difference in single-particle orbital occupations, $\delta n \equiv \langle \hat n_0 \rangle - \langle \hat n_1 \rangle$.
When the energy separation is large we expect all electrons to form a $\nu = \frac{1}{2}$ state in the lowest-energy Landau level, as the interactions should not be able to significantly hybridize the single-particle orbitals.
In that case, we would have $\langle \hat n_0 \rangle = \frac{1}{2}$ and  $\langle \hat n_1 \rangle = 0$, or $\delta n = \frac{1}{2}$. 
As the energy separation gets smaller, the system could valley polarize, i.e. pick a coherent, equal-amplitude superposition of the Landau levels.
Given the symmetry of the problem, the exact ground state should be a ``cat state'' superposition of the two states in which all electrons are in the same valley; 
but such a state would require a large entanglement to simulate effectively. Since DMRG only allows for limited entanglement, it spontaneously breaks the symmetry and leads to a valley-polarized ground state\cite{Zaletel2015}.
Such a state would have $\langle \hat n_0 \rangle = \langle \hat n_1 \rangle = \frac{1}{4}$, hence $\delta n = 0$.

\begin{figure}
\centering
\includegraphics[width = \columnwidth]{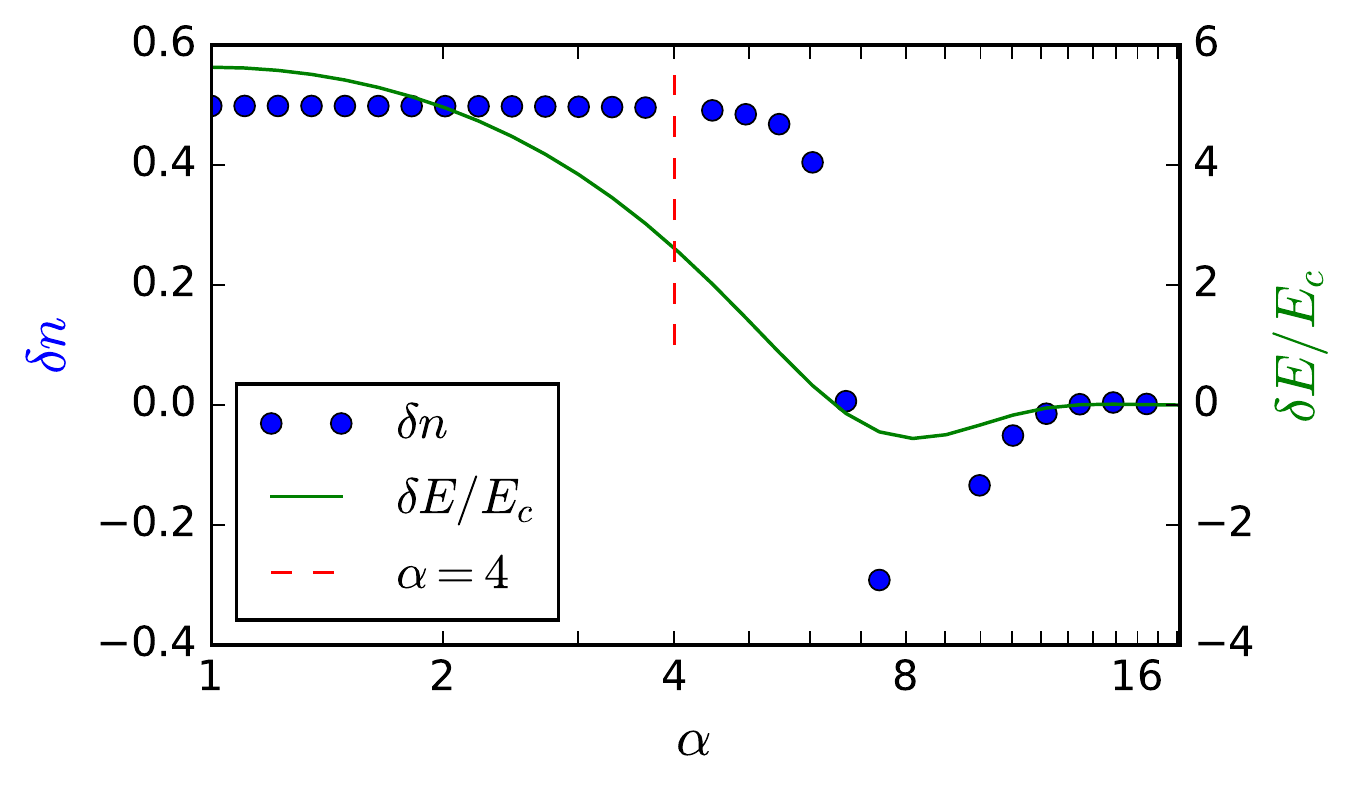}
\caption{Difference in orbital densities ($\delta n \equiv \langle \hat n_0 \rangle - \langle \hat n_1 \rangle$) as a function of $\alpha$ for the ground state of the Hamiltonian \eqref{H_anis} with Coulomb $e$-$e$ interactions at filling $\nu = \frac{1}{2}$ (blue dots). 
The values are computed via infinite DMRG on a cylinder of circumference $13 \ell_B$, with bond dimension $\chi \sim 3000$.
Magnetic field and electron density are the same as in Figure~\ref{fig:energies}.
The strength of $e$-$e$ interactions is taken to be $e^2 / \ell_B = E_0 \sim 2.4 E_c $.
Also shown is the inter-Landau-level gap $\delta E$ in units of cyclotron energy (red line).
$\delta n = 1/2$ implies that all the electrons are in one valley-symmetric Landau level, while $\delta n = 0$ is compatible with valley polarization.
}
\label{fig:dmrg}
\end{figure}

The results of our numerical calculation of $\delta n$ can be seen in Fig.~\ref{fig:dmrg}.
They indicate that the system does {\it not} valley-polarize until the interaction strength $e^2 / \ell_B $ is several times larger than the inter-Landau-level gap $\delta E$, 
{\it viz.} $\kappa \approx 5$.
Nearly all electrons populate the ``symmetric'' Landau level up to $\alpha \approx 6$.
After the Landau levels cross at $\alpha \simeq 6.45$, we see that a small energy difference ($\approx 0.4E_c$) in favor of the ``anti-symmetric'' Landau level is enough to make $\delta n$ significantly negative.
Valley polarization is thus only possible in the immediate vicinity of the crossing ($\alpha \simeq 6.45$) and beyond $\alpha \approx 12$.
We remark that our choice of Landau level mixing parameter $\kappa$ at the isotropic point, $\kappa \approx 0.5$, is conservative; the desirable scenario often assumed in studies of quantum Hall physics is that of negligible Landau level mixing, $\kappa \ll 1$.
Such a choice of parameters would further hinder valley polarization and reinforce our conclusion.

By studying the density-density correlator of states with $\delta n \simeq 1/2$, 
i.e. for $\alpha < 6$, we are able to map the Fermi contour of CFs with a method that has been detailed elsewhere\cite{Geraedts, Ippoliti1}.
This shows that the composite Fermi contour goes from being a circle at the isotropic point (in agreement with the analysis of Section~\ref{sec:exact}) to an ellipse which becomes more and more elongated as the parameter $\alpha$ is increased\footnote{The ratio of longest to shortest Fermi wavevector varies between 1 and $\simeq 2$ in the available range of $\alpha$.}.


Thus, we again find a lack of correspondence between the Fermi contours of the zero-field electrons and the CFs:
while the electron Fermi sea is made of two well-separated pockets like in Figure~\ref{fig:beans}, the CFL has a Fermi sea consisting of a {\it single} connected component, generally elliptical, but adiabatically connected to the circular one at $\alpha=1$.
In the same situation, but with total filling $\nu=1$ (i.e. 1/2 per Fermi pocket), the system simply forms a $\nu=1$ integer quantum Hall state with the generalized lowest Landau level orbitals.

If we take $\alpha$ to be very large, to the point that the inter-Landau-level gap becomes much smaller than the interaction strength, then the system is described by a bilayer with linearly-independent form factors.
At total filling $\nu=1$, such a system is expected to spontaneously valley-polarize \cite{Sodemann2017}, again forming an integer quantum Hall state.
At total filling $\nu=\frac12$, the system is expected to have a rich phase diagram including potentially a Halperin $331$ state, a Moore-Read state, two $\nu=1/4$ CF Fermi contours (one in each valley), or a valley-polarized $\nu=1/2$ CF Fermi sea (the history of such systems is reviewed in Ref.\cite{Peterson2010}).
Either way, the two zero-field electron pockets are not mapped into two $\nu = \frac{1}{2}$ CF pockets.
This conclusion is in agreement with experimental observations on electrons in GaAs quantum wells in parallel magnetic fields \cite{Kamburov2014}, where the CF Fermi contour is found to remain connected up to the highest achievable values of the parallel fields, while the electron Fermi sea is split into apparently well-separated pockets.

Finally, we comment on the dependence of our results on magnetic field.
In general, we expect two Fermi pockets to lead to well-defined quantum Hall valleys if the separation between them is large compared to $\ell_B^{-1}$, the typical spread of a Landau orbital in momentum space.
At the same time, following Eq.~\eqref{eq:cyclotron}, we know that the inverse magnetic length is 
\begin{equation} 
\ell_B^{-1} = \sqrt{\frac{\mathcal A(\epsilon_F)}{2\pi \nu} } \;,
\label{eq:ellB}
\end{equation}
so that the valley approximation holds well if the separation between the pockets is large enough relative to their typical linear size.
Moreover, Eq.~\eqref{eq:ellB} implies that the same set of pockets may or may not lead to quantum Hall valleys depending on filling, with larger $\nu$ being more likely to have well-defined valleys. For small chemical potential $\epsilon_F$ near the bottom of the band, we have small, widely separated pockets at zero field, and correspondingly in the high field (say, filling $\nu = \frac{1}{2}$), there are well-defined quantum Hall valleys.
Increasing $\epsilon_F$ causes the Fermi pockets to get closer and eventually merge, at which point there is no notion of valleys.
When exactly this transition occurs depends on the value of the Landau level mixing parameter $\kappa$, i.e. the ratio of interaction strength to inter-Landau-level gap. 
A reasonable estimator is the value $\alphacrit$ at which the two Landau levels cross for the first time. 
We computed $\alphacrit$ for a range of magnetic field values $B$ and found an approximate linear dependence of the former on the latter. This confirms that larger electron densities (larger $\epsilon_F$), which require stronger magnetic fields to get into the lowest Landau level quantum Hall regime, tend to oppose the onset of a many-valley regime and instead favor the population of a single non-degenerate, valley-symmetric, lowest Landau level.

\begin{figure}
\centering
\includegraphics[width = 0.9\columnwidth]{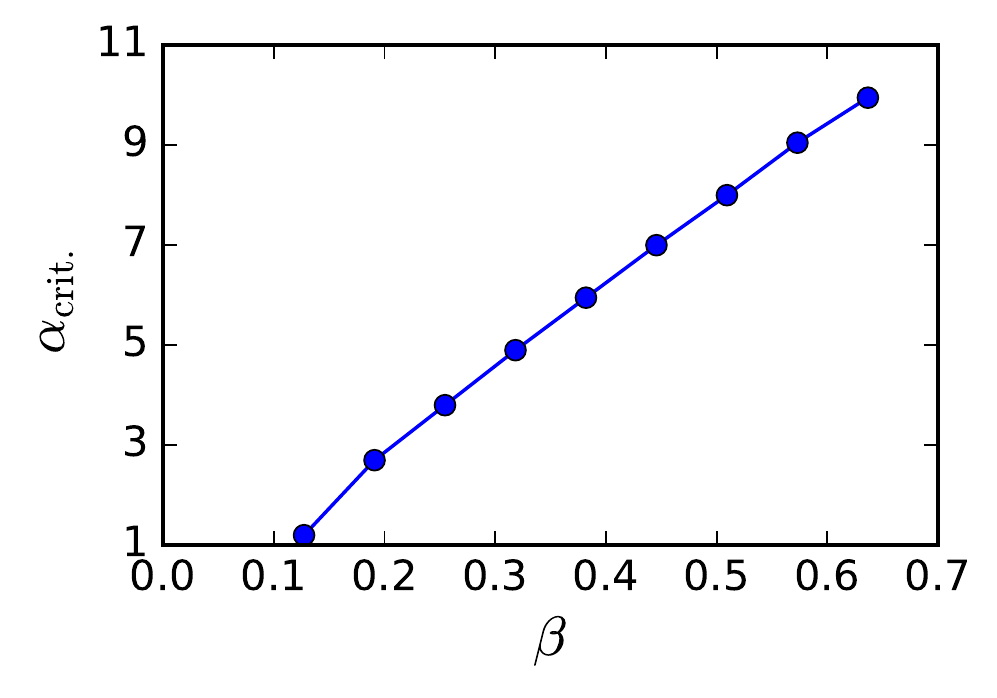}
\caption{Dependence of $\alphacrit$ (value of anisotropy when the first Landau level crossing occurs) on the magnetic field, expressed through the dimensionless ratio $\EE = E_c / E_0 = B (\hbar e / mE_0)$. 
For each value of $\beta$, we fix magnetic field and density (at filling $\nu = \frac{1}{2}$) and increase $\alpha$ starting from 1 until we encounter the first crossing.
For example, Figure~\ref{fig:energies} correspond to $\EE = 0.41$ and has the first crossing at $\alpha \simeq 6.45$.
\label{fig:alpha_crit}}
\end{figure}

\section{Conclusions}

We have investigated the connection between the Fermi contours of zero-field electrons/holes and their high-field, $\nu = \frac{1}{2}$ composite fermion counterparts. 
We divided our analysis into two cases of interest: 
that of a rotationally symmetric zero-field dispersion, potentially giving rise to a Fermi sea consisting of one or more annuli, 
and that of an explicitly symmetry-breaking dispersion, giving rise to multiple disconnected pockets. 

In both cases, we find that these shapes are not straightforwardly transferred to the composite Fermions.
An exact analytical argument allows us to prove that the CFL is {\it completely insensitive} to the rotationally-symmetric distortion, so that its Fermi contour remains a circle, regardless of whether the system at zero field develops a disconnected Fermi sea.
A similar conclusion extends to the many-valley case, where the CFL either has a single connected Fermi sea, or transitions into different phases, depending on the separation of the pockets and the magnetic field.

Finally, an interesting consequence of our analysis in the rotationally symmetric case is the possibility of using the band structure to induce a rearranging of the Landau levels, so that {\it any} Landau orbital can be made to have the lowest energy. 
This gives rise to the interesting possibility of observing higher-Landau-level physics, such as stripe or bubble phases, in the high-field regime.

\acknowledgments

This work was supported by Department of Energy BES Grant DE-SC0002140. DMRG calculations were performed using the TenPy library written by R. Mong and M. Zaletel.

\appendix

\section{Lack of correspondence between topology of electron and CF Fermi seas}
\label{app:rings}

In this Appendix we show an example of circularly symmetric dispersion where the electron Fermi sea can be made to have any number of disconnected components by tuning the Fermi energy, while the CFs always have the same circular Fermi sea.
This complements and generalized the result of Section~\ref{sec:exact} about the annular Fermi sea obtained from a quartic dispersion.

\begin{figure}
\includegraphics[width=0.9\columnwidth]{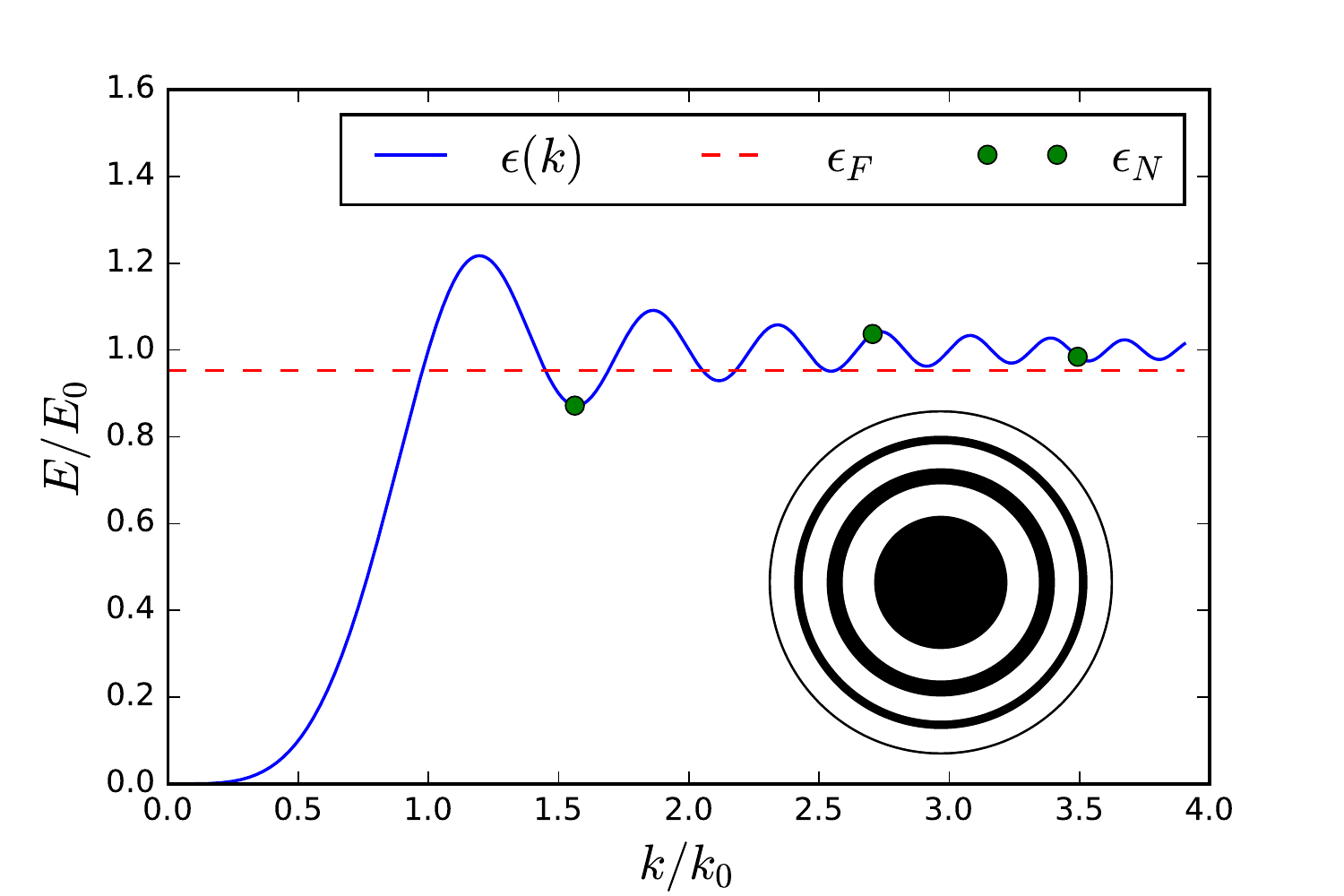}
\caption{Dispersion of Hamiltonian \eqref{eq:Ham_series} (solid line) with Fermi energy $\epsilon_F = 0.953E_0$ (dashed line) and energy eigenvalues for the lowest 3 Landau levels (dots).
Inset: shape of the corresponding zero-field Fermi sea, with a circle and 3 rings.}
\label{fig:target}
\end{figure}

Consider the function $f(x) = 1 - \sin(2\pi x) / 2\pi x$, i.e. the one-electron isotropic dispersion 
\begin{equation}
\epsilon(\vec k)
= E_0 \left( 1 - \frac{\sin(\pi  k^2/k_0^2)}{\pi  k^2/k_0^2} \right) \;.
\label{eq:Ham_series}
\end{equation}
This dispersion asymptotically approaches $E_0$ as $k\to \infty$, with infinitely many minima that get progressively shallower (we neglect effects of the finite Brillouin zone size here). 
Thus, as the Fermi energy approaches $E_0$ from below, one gets a Fermi sea consisting of arbitrarily many rings.
Specifically, $\mathcal A(\epsilon_F)$ (the area of the Fermi sea in $k$-space at chemical potential $\epsilon_F$) is a continuous, monotonically increasing function that goes from $\mathcal A = 0$ at $\epsilon_F \leq 0$ to $+\infty$ at $\epsilon_F \geq E_0$ (where the Fermi sea would include all but a finite area of momentum space). 
It diverges like $|\epsilon_F-E_0 |^{-1}$ as $\epsilon_F \to E_0^-$.
Now, since $\mathcal A(\epsilon_F)$ spans all positive real numbers as $\epsilon_F \to E_0^-$, 
a value of $\epsilon_F$ can be chosen such that the lowest Landau level energy $\epsilon_0$ (in the notation of Eq.~\eqref{energy}) falls in a local minimum of $f(x)$, 
which occurs near each integer value of $x$.

Given one such minimum $x_n \simeq n$, $f$ is such that $f(x) > f(x_n)$ for all values $x>x_n$.
Therefore, all higher Landau levels $N>0$ are guaranteed to have $\epsilon_N > \epsilon_0$, so that $\text{argmin}_N(\epsilon_N) = 0 $ and the ground state is a $\nu = \frac{1}{2}$ CFL made from the familiar $N = 0$ Landau orbitals; hence its Fermi sea is circular.
The (arbitrary) choice of minimum $n$ also determines the number of rings in the corresponding zero-field Fermi sea.
It is straightforward to prove that this number scales as $O(n)$ and, in particular, that it can be made arbitrarily large.
An example is illustrated in Fig.~\ref{fig:target}, where $n=1$ gives rise to a circle and 3 annuli.

\bibliography{annulus}

\end{document}